\documentclass[12pt]{iopart}
\usepackage{iopams}

\expandafter\let\csname equation*\endcsname\relax
\expandafter\let\csname endequation*\endcsname\relax

\usepackage{graphicx,latexsym,color,dcolumn,bm,epsfig}
\usepackage{amsmath}
\usepackage{amssymb}
\usepackage{amsfonts}
\usepackage[latin1]{inputenc}
\usepackage{subfigure}
\usepackage{cite}

\def\bbm[#1]{\mbox{\boldmath $#1$}}
\newcommand{\ket}[1]{\displaystyle{|#1\rangle}}

\begin{document}

\title{High-bandwidth quantum memory protocol for storing single photons in rare-earth doped crystals}
\author{Valentina Caprara Vivoli, Nicolas Sangouard, Mikael Afzelius, Nicolas Gisin}
\address{Group of Applied Physics, University of Geneva, CH-1211 Geneva 4, Switzerland}

\date{\today}

\begin{abstract}
We present a detailed analysis of a high-bandwidth quantum memory protocol for storing single photons in a rare-earth-ion doped crystal. The basic idea is to benefit from a coherent free-induced decay type re-emission which occurs naturally when a photon with a broadband spectrum is absorbed by a narrow atomic transition in an optically dense ensemble. This allows for a high-bandwidth memory for realistic material parameters. Long storage time and on-demand readout are obtained by means of spin states in a lambda-type configuration, through the transfer of the optical coherence to a spin coherence (so-called spin-wave storage). We give explicit formulas and show numerical results which make it possible to gain insight into the dependence of the memory efficiency on the optical depth and on the width and the shape of stored photons. We present a feasibility study in rare-earth doped crystals and show that high efficiencies and high bandwidth can be obtained with realistic parameters. High-bandwidth memories using spin-wave storage offers the possibility of very high time-bandwidth products, which is important for experiment where high repetition rates are needed.
\end{abstract}

\submitto{\NJP}
\maketitle

\section{Introduction}

Quantum memories are devices capable of storing single photon states in a stationary medium \cite{Lvovsky2009,Hammerer2010,Simon2010,Tittel2010b,Bussieres10}, which are essential building blocks of emerging quantum technologies. Quantum memories generally allows one to synchronize independent and probabilistic quantum processes, both for quantum communication schemes based on quantum repeaters \cite{Briegel1998,Duan2001,Sangouard2011} and quantum computing/simulation based on linear optics \cite{Knill2001,Kok2007}. Many different quantum memory schemes have been proposed, based on, for instance, electromagnetically induced transparency (EIT), Raman interactions, gradient echo memories (GEM) or atomic frequency combs (AFCs). Some of these have been demonstrated both in atomic vapours and solid-state ensembles, such as EIT and GEM, while others like Raman (in vapours) and AFC memories (in solid-state crystals) have so far only been demonstrated in a particular medium. For a comprehensive review of these techniques we refer to the review articles \cite{Lvovsky2009,Hammerer2010,Simon2010,Tittel2010b,Bussieres10} and references therein.

We focus on implementation of quantum memories in solid-state crystals, more precisely crystals doped with rare-earth (RE) ions, see the review \cite{Tittel2010b}. These exhibit optical inhomogeneous broadening typically of the order of 100 MHz to 10 GHz, induced by crystals strains or interactions with other impurities. At low temperatures, typically below 4 K, the homogeneous linewidth can be in the range of 1 kHz to 1 MHz, depending on the experimental conditions (temperature, external magnetic field value, doping concentration, etc.). All so-far proposed quantum memory schemes for RE crystals have focused on using the large number of spectral channels. Consequently these are generally based on control of the associated inhomogeneous dephasing, so-called photon echo schemes. Proposed methods to overcome this dephasing include controlled and reversible inhomogeneous broadening (CRIB) \cite{Nilsson2005,Kraus2006,Hetet2008}, atomic frequency combs (AFC) \cite{Afzelius2009a} or modified two-pulse photon echo schemes \cite{Damon2011,McAuslan2011}. To this date, light storage at the single photon level has mainly been demonstrated using the AFC scheme, eg. \cite{Riedmatten2008,Clausen2011,Saglamyurek2011}, but also with the CRIB scheme \cite{Hedges2010,Lauritzen2010}. The bandwidth of these memories will likely always be limited to the MHz regime, the major constraints being either limitations in the Rabi frequencies of the optical control fields or constraints in the available bandwidth over which the optical pumping can work. This provides a motivation to investigate alternative ideas for achieving very high bandwidth memories that could function with high repetition rates.

Here, we propose and analyse a scheme that could provide a solid-state quantum memory with very large bandwidth, exceeding 10 GHz. This memory would use the entire optical inhomogeneous broadening and the mechanism used allows one to store optical pulses with bandwidths considerably larger than the inhomogeneous bandwidth of the system. This is inspired by schemes proposed for homogeneous ensembles \cite{Gorshkov2007,Gorshkov2007c,Gorshkov2007d}, where a short pulse can be completely absorbed by the medium by properly shaping the temporal evolution of the pulse. To achieve long-term storage we combine this idea with so-called spin-wave storage, meaning that optical $\pi$-pulses are used to convert the optical coherence into a spin coherence to exploit the long spin coherence times and allow on-demand read out.

We believe that this high-bandwidth memory will likely find other, complementary use, as compared to, for instance, AFC quantum memories, where quantum repeaters is the main application of such multimode quantum memories. High-bandwidth, high-rate memories are likely to be useful for smaller scales \cite{Reim2010,Reim2011,England2012}, such as high-rate table-top experiments, in the context of generating high photon number quantum states \cite{Nunn2013} or more generally in quantum computing/simulation based on linear optics \cite{Knill2001,Kok2007}. We also believe that the basic scheme can be adapted in order to create a microwave-to-optical interface using RE crystals coupled to superconducting waveguides. Several groups actively work towards this goal, some based on RE doped solids \cite{Bushev2011,Staudt2012,Probst2013}. In short a single excitation induced by the absorption of a single microwave photon can in principle be read out as an optical photon. The scheme analyzed here is relevant for these ideas, since we are using and manipulating the entire optical inhomogeneous broadening, which will be required also for a microwave-to-optical interface.

The article is organised in the following way. In section \ref{Principle} we present the basic idea of the protocol. In section \ref{Analyticalformulas} we derive analytical expressions for the output field given a particular input field shape. In section \ref{optimizedefficiency} we study the optimization of the efficiency for a given optical depth. Finally in section \ref{implementation} we discuss the prospect of implementing the scheme in RE doped crystals.

\section{The principle\label{Principle}}
 \begin{figure}[h!]
 \
\subfigure{\includegraphics[width=0.48\textwidth]{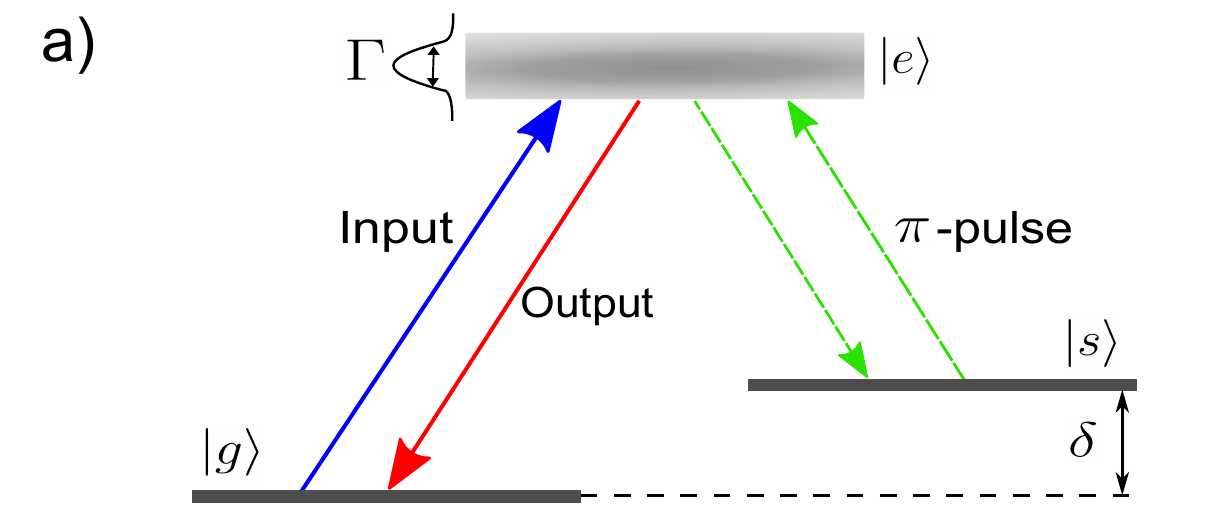}\label{AC}}
\hspace{1.5cm}
\subfigure{\includegraphics[width=0.48\textwidth]{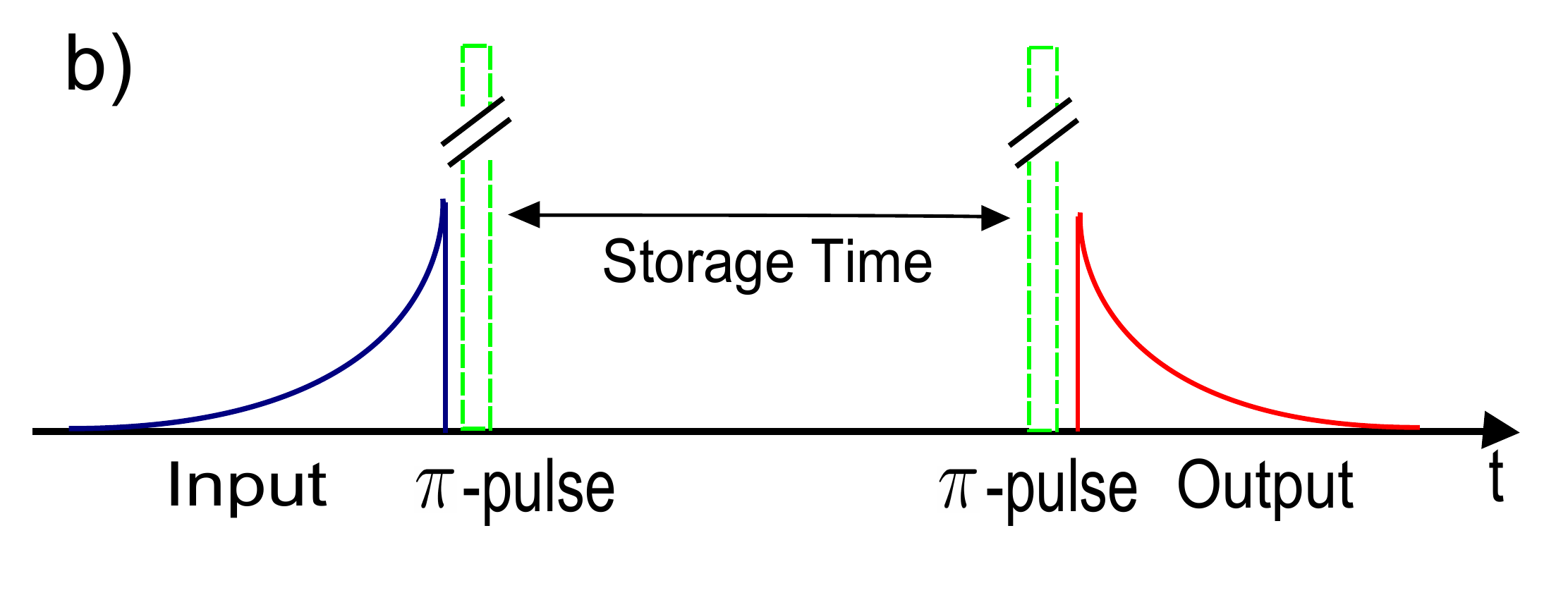}\label{SP}}
\caption{a) The principle of the proposed memory protocol. The atomic spectrum is composed by three energetic levels. An inhomogenously broadening is associated to the atomic transition $\ket{g}-\ket{e}$. The $\ket{s}$ is an auxiliary state used to avoid the dephasing of the atomic excitation during the storage time. b) Pulse sequence of the memory protocol. The input is completely absorbed by the ensemble during a lapse of time of the order of $T\ll 1/\Gamma$. Since the dephasing time is of the order of $1/\Gamma$, the system doesn't dephase during the absorption process. The excitation is transferred in the auxiliary state through a $\pi$-pulse so that there is not dephasing neither during the storage time. Another $\pi$-pulse transfers again the  population in the excited state bringing to the re-emission process.}
\end{figure}

The storage material of length $L$ is composed of a set of $N$ atoms, each with an excited state $\ket{e}$ and two lower states $\ket{g}$-$\ket{s}$, see Fig. \ref{AC}. We will here consider so-called Kramers ions, with uneven number of electrons (eg. Er$^{3+}$, Nd$^{3+}$ or Yb$^{3+}$), which usually have doubly degenerate ground-state levels with a magnetic sensitivity of the order of the Bohr magneton (hence proportional to 14 GHz/Tesla). These Zeeman levels can thus easily be split in order to have optically resolved transitions between all magnetic sub-levels \cite{Afzelius2010b}. These ions are also being investigated for interfacing with superconducting quantum circuits \cite{Bushev2011,Staudt2012,Probst2013} since with a reasonable magnetic field one can obtain transitions at the operating frequency of around 5 GHz.

We assume that the optical transitions $\ket{e}$ -- $\ket{g}$ and $\ket{e}$ -- $\ket{s}$ are inhomogeneously broadened, in the GHz regime, whereas the spin transition $\ket{g}$ -- $\ket{s}$ is assumed to have a considerably less inhomogeneous broadening. Recent measurements in Er$^{3+}$:Y$_2$SiO$_5$ have shown spin linewidths between 10 and 100 MHz \cite{Bushev2011,Staudt2012,Probst2013}. Furthermore, spin echo techniques can be used so that a coherence between $\ket{g}$ and $\ket{s}$ can be preserved during time intervals comparable to the homogeneous coherence lifetime, which can approach 100 $\mu$s at cryogenic temperatures \cite{Bertaina2007}. \\

We consider an input field having many spectral components (centered at $\omega_0$). Specifically, its spectral bandwidth $1/T$ is assumed to be much larger than the bandwidth of the inhomogenenous spectrum $\Gamma$ ($\Gamma$, $1/T$ are defined as the half width half maximum in the following). This does not prevent the complete absorption of the input pulse if the optical depth of the atomic medium $\alpha L$ is high enough
\cite{Gorshkov2007c,Gorshkov2007d}. Indeed, even though the number of atoms in the tails of the atomic distribution is smaller than the number of atoms at the central frequency, they can efficiently absorb the side frequencies of the input field if the peak absorption $\alpha L$ is high enough. Close to complete absorption is reached provided that $\alpha L \Gamma T\gg 1.$\\

Further consider the quantum regime where the input pulse contains a single photon. The atoms are assumed to be prepared initially in $|g\rangle.$ Right after the absorption, say at time $t=0,$ the photon is mapped into a collective atomic excitation which is described by
\begin{equation}
\label{collective}
\sum_{i=1}^{N} c_i e^{i k z_i-i \Delta_i t} |g_1\hdots e_i\hdots g_N \rangle.
\end{equation}
The sum runs over the $N$ atoms. $k$ is the wave number of the input photon. $c_i$ stands for the probability amplitude for having an absorption by the atom $i.$ A priori, $c_i$ depends on both the position $z_i$ of the atom $i$ and on the frequency detuning $\Delta_i=\omega_i-\omega_0$ between the atom $i$ and the photon. Importantly, in the regime that we consider where $\Gamma T \ll 1,$ the argument of the exponential term does not depend on $\Delta_i.$ Indeed, the absorption process has a duration that is essentially given by the input pulse duration $T$ which is much shorter than the dephasing time $\Gamma^{-1}.$ All phase factors of the form $e^{i \Delta_i t}$ thus reduces to unity (they are bounded by $e^{i\Gamma T} \approx 1$). Therefore, the components of the collective atomic excitation (\ref{collective}) are in phase after the absorption, leading to a coherent free-induced decay type re-emission in the forward direction, i.e. the re-emission of a light field propagating in the same direction as the input photon. \\

This describes a basic quantum memory for which the emission time is not controlled. For an on-demand re-emission, the single collective atomic excitation is transferred to the spin transition by means of a $\pi$-pulse resonant with the transition $|e\rangle$-$|s\rangle$ with a temporal duration shorter than the one of the input pulse. The resulting collective spin excitation allows one to achieve long storage time, limited by the coherence lifetime of the spin transition only. It can then be readout by applying a second $\pi$-pulse. As we will show in the following, the readout efficiency is higher if the output pulse propagates in the backward direction. This can be done by choosing two $\pi$-pulses propagating in opposite directions \cite{Nilsson2005}. Both forward and backward readouts are studied in detail in the next section where we give explicit formulas for the efficiency of the storage and retrieval steps.\\

\section{Analytical formulas for efficiency\label{Analyticalformulas}}
In this section, we aim at deriving analytical formulas for the efficiency of absorption and re-emission processes. We work with a one-dimensional light propagation model in which the electromagnetic field is described by slowly varying forward and backward modes $E_f(z,t)$ and $E_b(z,t)$ respectively. Similarly, the atomic coherence $\sigma_{ge}(z,t,\Delta)$ associated to the frequency detuning $\Delta$ is decomposed into forward and backward components $\sigma_f(z,t,\Delta)$ and $\sigma_b(z,t,\Delta)$. (We now consider the continuous limit where $\Delta_i \rightarrow \Delta.$) Assuming that most of the population remains in the ground state during the storage and retrieval processes, we can show that the light field and the atomic coherence satisfy the following equations (see for details \cite{Sangouard2007},\cite{Afzelius2009a}):
\begin{equation}
 \frac{\partial}{\partial z}E_f(z,t)=\frac{i g^2 N}{c}\int_{-\infty}^{\infty}d \Delta n(\Delta)\sigma_f(z,t;\Delta)\label{equEf}
\end{equation}
\begin{equation}
 \frac{\partial}{\partial t}\sigma_f(z,t;\Delta)=-i \Delta \sigma_{f}(z,t;\Delta)+i \wp E_f(z,t),\label{equsf}
\end{equation}
\begin{equation}
 \frac{\partial}{\partial z}E_b(z,t)=-\frac{i g^2 N}{c}\int_{-\infty}^{\infty}d\Delta n(\Delta) \sigma_b(z,t;\Delta)\label{equEb}
\end{equation}
\begin{equation}
 \frac{\partial}{\partial t}\sigma_{b}(z,t;\Delta)=-i \Delta \sigma_{b}(z,t;\Delta)+i \wp E_b(z,t).\label{equsb}
\end{equation}
where $\wp$ is the dipole moment of the transition $\ket{g}-\ket{e}$ and $g^2=\wp \frac{ \omega_0}{2\epsilon_0 V}$ is the single-photon single atom coupling constant and $\Delta$ is measured in rad$\cdot$Hz. $n(\Delta)$ is the atomic spectral distribution which satisfies the normalization condition $\int_{-\infty}^{+\infty}d\Delta n(\Delta)=1$. In this section, $n(\Delta)$ is assumed to be a Lorentzian (with full width at half maximum (FWHM) $2\Gamma$)
\begin{equation}
n(\Delta)=\frac{\Gamma}{\pi(\Gamma^2+\Delta^2)}.\label{AtomicDistribution}
\end{equation}
We also assume that at the entrance of the atomic ensemble the field is described by
\begin{equation}
\bigg\{
 \begin{array}{ccccc}
 E_f(0,t)&=&\sqrt{\frac{2}{T}}e^{t/T}&\,\,\,\,\,&t\le0\\
 E_f(0,t)&=&0&\,\,\,\,\,&t>0
 \end{array}
\label{equInCon}
\end{equation}
where the input intensity duration is $T/2$ and the prefactor is being chosen such that $\int_{-\infty}^{0}dt \lvert E_f(0,t)\rvert ^2=1$. The choice of an envelop that is null for positive times makes it easy to distinguish the input and the re-emitted parts of the light field. The choice of an exponentially varying pulse for negative times is natural as it matches the mode that is spontaneously and coherently emitted by an ensemble of atoms following a Lorentzian spectral distribution \cite{Gorshkov2007,Gorshkov2007c,Gorshkov2007d}, c.f. below for details. The assumptions (\ref{AtomicDistribution})-(\ref{equInCon}) allow one to find analytical expressions for the efficiency of the memory protocol in several regimes, as we shall see now.

\subsection{Absorption process}
Let us first focus on the absorption process. We consider an input field propagating in the forward direction $E_f(z,t)$ that couples to the atomic coherence $\sigma_f(z,t;\Delta)$ through Eqs. (\ref{equEf}) and (\ref{equsf}). We initially assume that the field maintains its initial shape as it propagates through the ensemble of atoms while its amplitude decreases
\begin{equation}
\label{guess}
E_f(z,t)=f(z)E_f(0,t).
\end{equation}
The formal solution of the Eq. (\ref{equsf}), which is given by
\begin{equation}
 \sigma_f(z,t;\Delta)=i \wp \int_{-\infty}^{t}dt' e^{-i \Delta(t-t')}E_f(z,t').
 \end{equation}
then reduces to
\begin{equation}
\sigma_f(z,t;\Delta)=\frac{i \wp T}{1+i\Delta T}E_f(z,t)\label{solsf}
\end{equation}
for the specific choice (\ref{equInCon}). Inserting Eq. (\ref{solsf}) in Eq. (\ref{equEf}) we find
\begin{equation}
  \frac{\partial}{\partial z}E_f(z,t)=-\frac{\alpha\Gamma T}{2} E_f(z,t)\int_{-\infty}^{\infty} d\Delta n(\Delta)\frac{1}{1+i\Delta T}
\end{equation}
with $\alpha =\frac{2\wp g^2N}{c\Gamma}.$
Further assuming that the atomic spectral distribution is given by Eq. (\ref{AtomicDistribution}), we obtain
\begin{equation}
 \frac{\partial}{\partial z}E_f(z,t)=-\frac{\alpha \Gamma T}{2\left(1+\Gamma T\right)} E_f(z,t)
\end{equation}
whose solution
\begin{equation}
E_f(z,t)=e^{-\alpha  z\frac{\Gamma T}{2(1+\Gamma T)}}E_f(0,t)
\end{equation}
validates our initial assumption (\ref{guess}). The efficiency of the absorption process which is defined as $\eta_{abs}=1-\frac{\int_{-\infty}^{0}dt \vert E_f(L,t)\vert^2}{\int_{-\infty}^{0}dt \vert E_f(0,t)\vert^2},$ is thus given by
\begin{equation}
\eta_{abs}=1-e^{-\frac{\alpha L \Gamma T }{1+\Gamma T}}.\label{Effab}
\end{equation}
This expression is valid for any value of $\Gamma T.$ Specifically, if the input pulse spectrum is well contained into the atomic spectral distribution $\Gamma T \gg 1,$ as in CRIB or AFC type memories, the last expression reminds us that  the transmission probability scales like the inverse of the exponential of the peak optical depth. It further shows that the absorption can be 100\% efficient, for any $\Gamma T,$ provided that the optical depth is large enough. We will see in the following that only the regime $\Gamma T\ll1$ leads to a constructive interference of individual emissions by means of a coherent free-induced decay type re-emission. In this case, the absorption process can be efficient if the optical depth averaged over the input pulse spectrum is large enough $\alpha L \Gamma T \gg 1.$

\subsection{Re-emission process}
\subsubsection{Backward emission \label{paremission}}
\begin{figure}[h]
\begin{center}
\includegraphics[width=260pt]{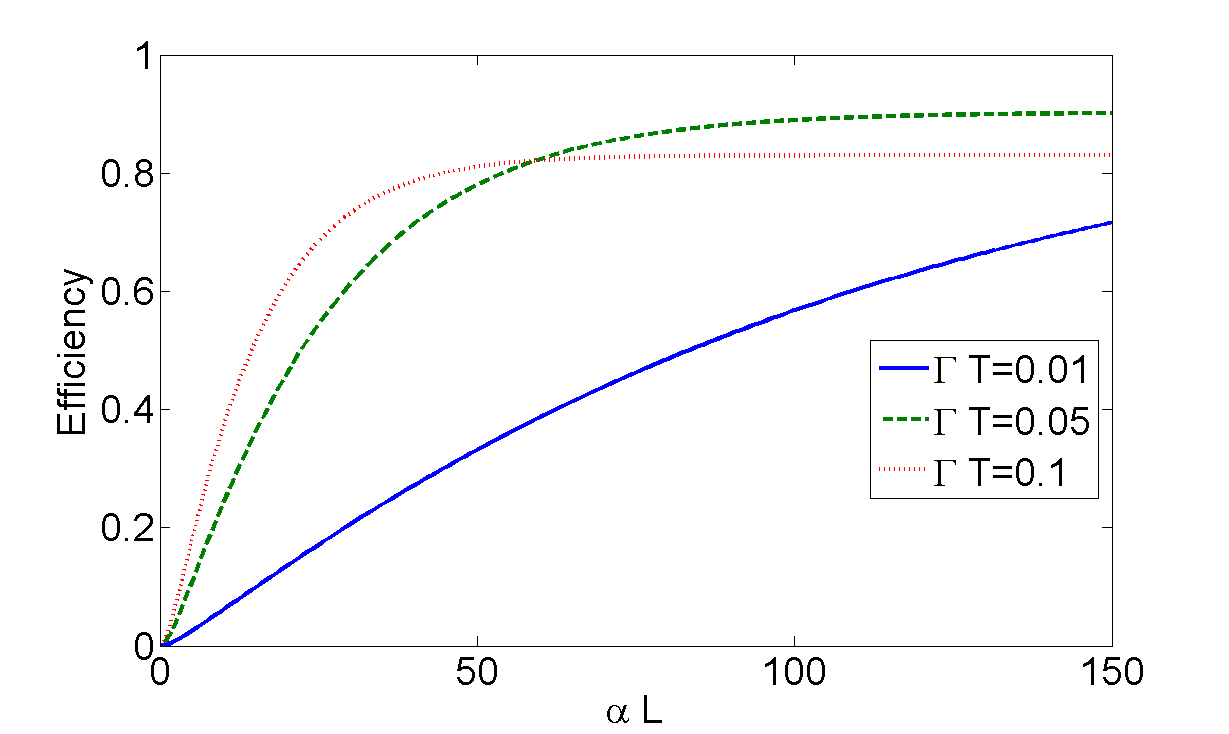}
\end{center}
\caption{Overall efficiency (storage and retrieval) of the backward memory protocol as a function of the optical depth $\alpha L$ for various values of $\Gamma T.$ The maximum efficiency is obtained in the limit $\alpha L \Gamma T\gg 1$ and can be computed from the Eq. (\ref{Effl}) ($\eta_{\text{back}}(\Gamma T=0.1)=0.83$,  $\eta_{\text{back}}(\Gamma T=0.05)=0.91$ and $\eta_{\text{back}}(\Gamma T=0.01)=0.98$.\label{fig2}}
\end{figure}

By applying the two $\pi$ pulses in opposite directions, the spatial phase pattern of the atomic polarization changes and the resulting collective atomic excitation couples to the backward mode of the electromagnetic field \cite{Nilsson2005}. Under the assumption that the $\pi$-pulses perfectly transfer the population back and forth between $\ket{e}$ and $\ket{s}$ and that they are short enough such that the dephasing of the collective atomic excitation is negligible, one can treat them as being applied essentially simultaneously at time $t=0$. The absorption process, thus, occured at negative times and re-emission process occurs at positive times. At $t=0,$ the atomic polarization is given by
\begin{equation}
\sigma_b(z,0^+;\Delta)=\sigma_f(z,0^-;\Delta)=\frac{i \wp T}{1+i\Delta T}E_f(z,0)\label{equInCon2}
\end{equation}
where $\sigma_{b(f)}\left(z,0^{\pm},\Delta\right)$ denotes the polarization after (before) the $\pi$ pulse sequence. Furthermore, the field $E_b$ is assumed to be zero initially
\begin{equation}
E_b(L,0)=0.\label{equInCon3}
\end{equation}
To know the efficiency of the retrieval process, one has to solve the Eqs. (\ref{equEb}) and (\ref{equsb}) with the initial conditions (\ref{equInCon2}) and (\ref{equInCon3}). From Eq. (\ref{equsb}), we can find the expression of the atomic coherence as a function of the light field
\begin{equation}
\label{coh_back}
\sigma_b(z,t;\Delta)=e^{-i\Delta t} \sigma_b(z,0^+;\Delta)-i \wp\! \int_0^t\!\! dt' E_b(z,t')e^{-i\Delta (t-t')}.
\end{equation}
Substituting $\sigma_b$ in Eq. (\ref{equEb}) and introducing the Fourier transform $E_b(z,\omega)=\frac{1}{\sqrt{2\pi}}\int_{-\infty}^{\infty}dt e^{i\omega t}E_b(z,t)$ we obtain at the position $z=0$
\begin{equation}
\begin{split}
E_b(0,\omega)&=-\frac{1}{1+2\Gamma T}\\
&\,\times E_f\left(0,\frac{-\omega}{1+2\Gamma T}\right)\left(1-e^{\frac{\alpha L \Gamma }{2(i\omega-\Gamma)}}e^{-\frac{\alpha L \Gamma T}{2(1+\Gamma T)}}\right).\label{SolEb}
\end{split}
\end{equation}
It clearly appears in this expression that the re-emitted field potentially undergoes distortion and attenuation effects. The first factor on the right-hand side (together with the renormalization factor of the frequency component of the forward field) stands for the irreversible dephasing of the atomic coherences during the absorption and re-emission steps and shows that input pulses for which $\Gamma T \gtrsim 1$ cannot be re-emitted efficiently. In the opposite regime where $\Gamma T \ll 1,$ the last exponential term tends to zero when $\alpha L \Gamma T \gg 1$ making negligible the distortion effect due to $e^{\frac{\alpha L \Gamma }{2(i\omega-\Gamma)}}.$ In this case, the storage reaches unity efficiency and the output pulse is simply the time reversal of the input. \\
To know precisely the memory efficiency $\eta_{\text{back}}$ as a function of the optical depth $\alpha L,$ we calculated the ratio between the energy of the output and input pulses, e.g. we numerically computed
\begin{equation}\begin{split}\label{eff_back_int}
&\eta_{\text{back}}=\frac{T}{2(1+2\Gamma T)^2}\\
&\times \int_{-\infty}^{+\infty}d\omega\left\lvert E_f\left(0,\frac{-\omega}{1+2\Gamma T}\right)\left(1-e^{\frac{\alpha \Gamma L}{2(i\omega-\Gamma)}-\frac{\alpha L \Gamma T}{2(1+\Gamma T)}}\right)\right\rvert ^2.
\end{split}\end{equation}
The results are shown in Fig. \ref{fig2} for various values of $\Gamma T$ and have been successfully compared with the ones obtained by solving numerically the Maxwell-Bloch equations \cite{Burr2004}. They show that the efficiency increases as $\Gamma T$ decreases, as expected, and tends to
\begin{equation}
\eta_{\text{back}}\rightarrow\frac{1}{1+2\Gamma T}\label{Effl}
\end{equation}
when $\alpha L \Gamma T\gg1.$

In the opposite regime where $\alpha L \Gamma T\ll1$, one can find an approximate solution by Taylor expansion, to first order, of the exponential functions in Eq. \ref{SolEb}. We find that the output pulse is not subject to any distortion and that the efficiency is well approximated by
\begin{equation}
\eta_{\text{back}}\approx \Gamma T\left(\frac{\alpha L}{2(1+\Gamma T)}\right)^2.
\end{equation}

In the regime where $\alpha L \Gamma T \approx 1,$ one cannot solve analytically the integral in (\ref{eff_back_int}) to give an explicit formula of the efficiency for any value of $\Gamma T.$ However, we will present in the next section an heuristic formula reproducing the memory efficiency when it is optimised over the input pulse duration. But before, we focus on the case where the output pulse propagates in the forward direction.\\

\subsubsection{Forward emission}

\begin{figure}[h]
\begin{center}
\includegraphics[width=260pt]{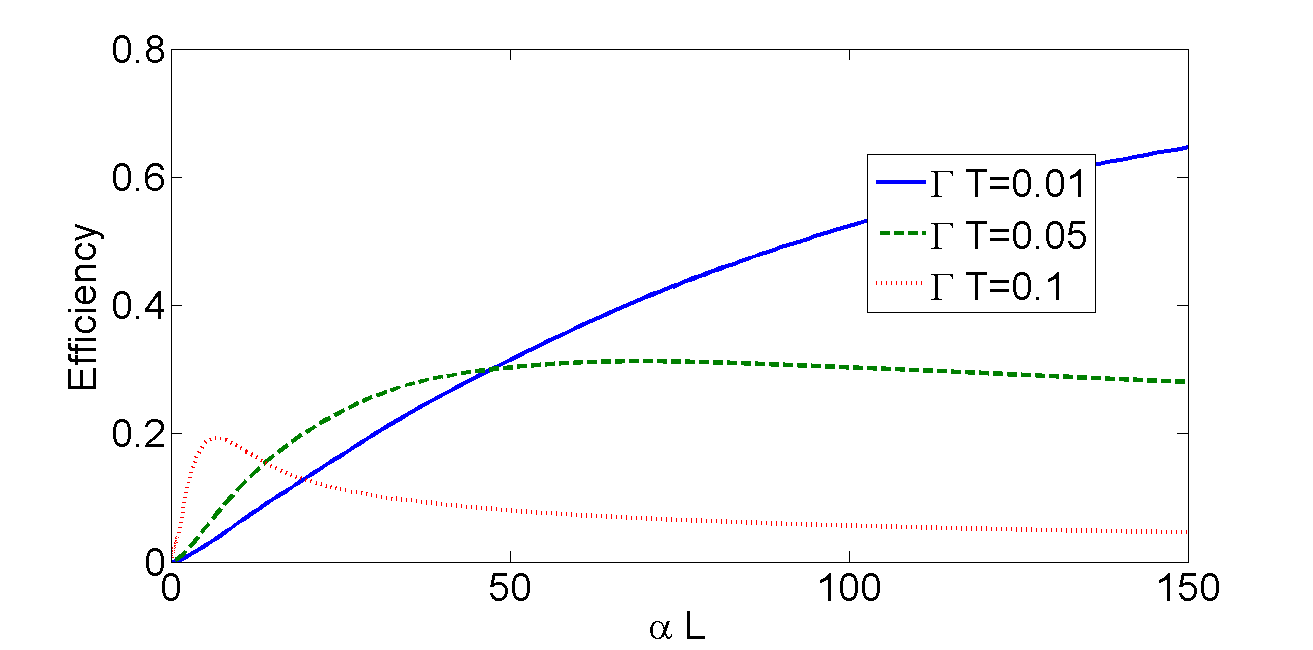}
\end{center}
\caption{Overall efficiency (storage and retrieval) of the forward memory protocol as a function of the optical depth $\alpha L$ for various values of $\Gamma T.$\label{fig3}}
\end{figure}

If the two control $\pi$-pulses propagate in the same direction, the collective atomic excitation couples to the forward mode of the electromagnetic field. In this case, the dynamics is obtained by solving the Eqs. (\ref{equEf}) and (\ref{equsf}). Following the previous line of thought, we can find the expression of the field re-emitted in the forward direction $\bar{E}_f(L,\omega)$ as a function of the input field $E_f(0,\omega)$
\begin{equation}
 \bar{E}_f(L,\omega)=e^{\frac{\alpha L \Gamma}{2(i\omega-\Gamma)}} E_f(0,\omega)\left(1-e^{-\frac{\alpha L \Gamma}{2(i\omega-\Gamma)}}e^{-\frac{\alpha L\Gamma T}{2(1+\Gamma T)}}\right).
\label{SolFor}
\end{equation}
Again, the light field potentially undergoes distortion and attenuation effects. Specifically, one can show that when integrated, the expression (\ref{SolFor}) leads to an overall efficiency that is identical to the one of the backward process in the limit $\alpha L \Gamma T\ll1.$ However, in the limit of high optical depths, the first exponential term strongly reduces the achievable efficiency, which tends to zero in the limit where $ \alpha L \rightarrow \infty.$ The main difference with the backward case is thus that the re-emission step comes potentially with re-absorption effects, leading to lower efficiencies in the limit of high optical depths. Fig. (\ref{fig3}) shows the forward efficiency as a function of the optical depth for various values of  $\Gamma T.$ The results are obtained by numerically integrating the expression (\ref{SolFor}), similarly to what has been done in the previous subsection. One can see that by choosing the input pulse spectrum larger than the atomic distribution spectrum, $\Gamma T \ll 1$, to reduce the atomic dephasing, one can obtain high efficiencies for certain optical depths. In particular, in the limit where $\Gamma T \rightarrow 0,$ the forward process can reach unity efficiency. This contrasts with the CRIB or AFC protocols where absorption effects during the re-emission limit the efficiency to 54\% \cite{Sangouard2007,Afzelius2009a}.\\

In the framework of a practical realization, it is important to know the maximum memory efficiency that can be obtained for a given optical depth. This is the aim of the next section where the memory efficiency is optimized over the duration and the shape of the input pulse for various inhomogenous profiles.\\

\section{Optimal efficiency\label{optimizedefficiency}}
\subsection{Optimization over the input pulse duration}
\subsubsection{Numerical results}

\begin{figure}[h]
\begin{center}
\includegraphics[width=250pt]{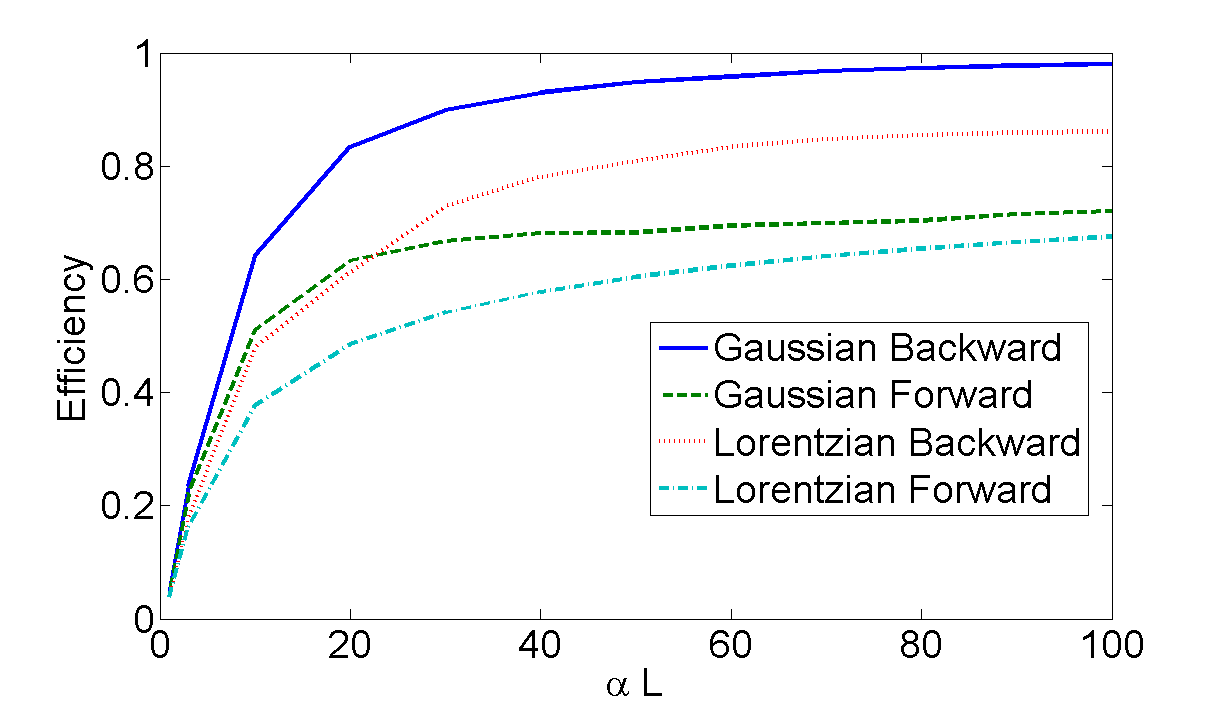}
\end{center}
\caption{Optimization of the memory efficiency over the input pulse duration $T$ as a function of the optical depth ($\alpha L$) assuming that its spectrum is Lorentzian. The forward and backward cases are compared for both Lorentzian and Gaussian inhomogeneous lines.
\label{fig4}}
\end{figure}

We here focus on the optimization of the memory efficiency over the input pulse duration. This is done numerically by solving the Maxwell-Bloch (MB) equations. Specifically, we start by assuming that the atomic spectral distribution is Lorentzian with fixed peak optical depth and we compute the overall efficiency (storage plus retrieval) for an input pulse with an exponential rising temporal shape. The computation is repeated for various input pulse durations until the efficiency is optimized. The results are presented in Fig. \ref{fig4} as a function of the optical depth. They show clearly that the efficiency in the forward case is lower than in the backward configuration due to unwanted re-absorption in the re-emission step. \\

We then compare the efficiency that is obtained for Gaussian inhomogeneous profile with FWHM $2\Gamma.$ In this case, $\alpha$ which is defined through the probability for an input photon (with a spectrum well contained into $\Gamma$) to be absorbed by a medium of length $L$ as $1-e^{-\alpha L},$ is given by $\alpha=\frac{g^2N}{2c\Gamma}.$ One sees in Fig. \ref{fig4} that the efficiency is the greatest in the case of the Gaussian atomic distribution. This is to be expected since for times small compared to $1/\Gamma$, a Gaussian distribution results in less dephasing than a Lorentzian distribution, as is evident by taking the Fourier transform of their frequency spectra.

\subsubsection{Analytical results}

\begin{figure}[h]
\begin{center}
\includegraphics[width=250pt]{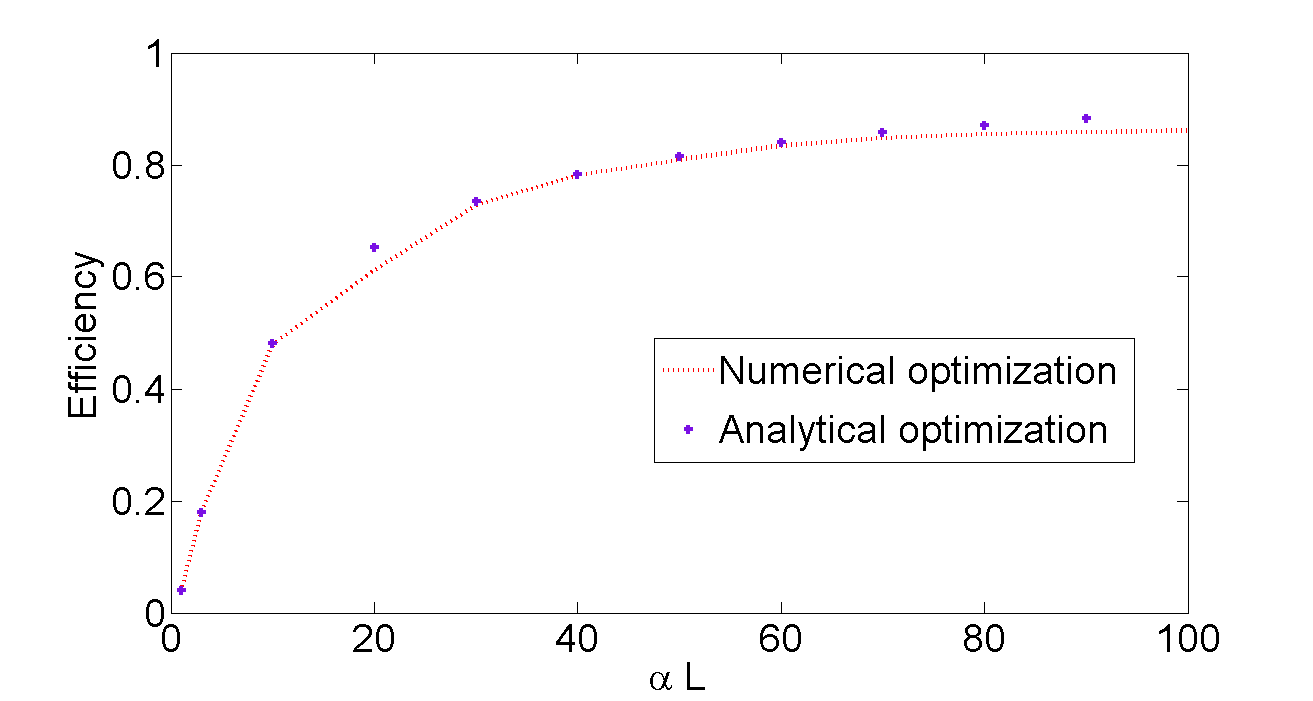}
\end{center}
\caption{Optimization of the memory efficiency over the input pulse duration $T$ as a function of the optical depth ($\alpha L$) assuming that its spectrum is Lorentzian (an exponentially rising temporal input shape) for a Lorentzian inhomogeneous line (backward case). This figure aims at comparing the results obtained numerically and the ones from the formula (\ref{equoptimizedeff}).
\label{fig5}}
\end{figure}

By studying the numerical results from the MB simulations, we observed that when storing a pulse with optimal duration, the efficiencies for absorption and re-emission processes are the same, for the backward process, provided that the potential dephasing of the atomic coherence is taken into account. We will here present a calculation that explicitly show this, for the optimized storage. To this end, instead of looking at the energy of the input that is not transmitted to quantify the absorption efficiency, as before, we now calculate the part of the input pulse that is coherently absorbed by the atoms, i.e. we estimate the coherent absorption through
\begin{equation}
\eta^*_{\text{abs}}=\frac{\int_0^L dz \frac{g^2 N}{\wp^2}\lvert \int_{-\infty}^{\infty}d\Delta n(\Delta)\sigma(z,0,\Delta)\rvert^2}{c\int_{-\infty}^0 dt \lvert E_f(0,t)\rvert^2}.
\end{equation}
(The prefactor $\frac{g^2 N}{\wp^2c}$ insures that the ratio is dimensionless.) In the case where the atomic inhomogeneous spectrum is Lorentzian and the input pulse has an exponentially increasing temporal shape, we find
\begin{equation}
\label{coh_abs}
\eta^*_{\text{abs}}=\frac{1}{1+\Gamma T}\left(1-e^{-\frac{\alpha \Gamma T  L}{1+\Gamma T}}\right).
\end{equation}
The only difference with the incoherent absorption (see Eq. \ref{Effab}) is the prefactor $1/\left(1+\Gamma T\right)$ that stands for the atomic dephasing which can occur during the absorption time if the input pulse duration $T$ is comparable or longer to the dephasing time $1/\Gamma$. Taking the square of the formula (\ref{coh_abs}) and optimizing numerically the result over $\Gamma T$ for a given optical depth, we find that the optimal pulse duration satisfies
\begin{equation}
\Gamma T_{\text{opt}} \approx \frac{1}{1+\frac{\alpha L}{4}}\label{fitGT}
\end{equation}
and leads to the following optimized overall efficiency
\begin{equation}
\eta_{\text{back}}^{\text{opt}} \approx \left(\frac{1+\frac{\alpha L}{4}}{2+\frac{\alpha L}{4}}\right)^2\left(1-e^{-\frac{\alpha L}{2+\frac{\alpha L}{4}}}\right)^2.\label{equoptimizedeff}
\end{equation}
Fig. \ref{fig5} shows that the formula (\ref{equoptimizedeff}) reproduces very well the results that are obtained numerically.

\subsection{Optimization over the input pulse shape}
Hitherto, the efficiency of the memory protocol has been optimized over the input pulse duration. We now look for the optimal input pulse shape, for a given inhomogeneous broadening. We here focus on the backward process in the case where the atomic absorption spectrum exhibits a Gaussian profile as inhomogeneous line usually does. For a fixed optical depth, we optimize the overall memory over the input pulse shape. To do that, we follow the procedure proposed in Ref. \cite{Gorshkov2007}, which is composed of three steps. First, an input pulse with an arbitrary shape is sent through the atomic medium. Second, the corresponding output pulse is time reversed and then sent again into the atomic ensemble. Third, the process is repeated until the output pulse shape is unchanged. This procedure converges towards the optimal input pulse, yielding the highest efficiency.

\begin{figure}[h]
\begin{center}
\includegraphics[width=250pt]{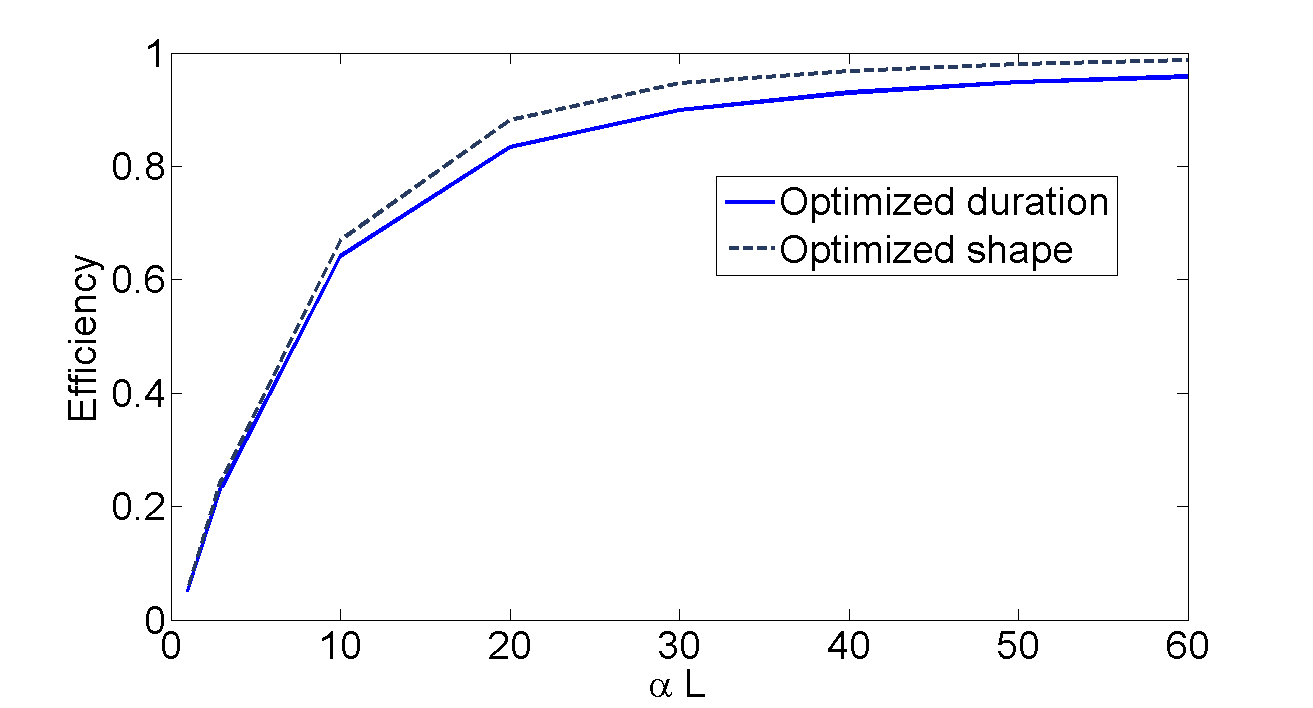}
\end{center}
\caption{Optimization of the memory efficiency as a function of the optical depth ($\alpha L$) over the input pulse duration and the input shape.The optimized shape temporal input for large optical dephts doesn't diverge from an exponential behaviour except that for small times.\label{fig6}}
\end{figure}

The numerical simulation has been performed for realistic optical depths, ranging from 0 to 60. By choosing an initial Lorentzian input spectrum, the procedure usually converges after three or four iterations. The results are shown in Fig. \ref{fig6} where they are compared with the ones obtained by optimizing only the pulse duration for a fixed Lorentzian shape (exponentially rising temporal shape), as in the previous subsection (compare to solid line in Fig. \ref{fig4}). For small optical depth $(\alpha L \leq 10),$ the Lorentzian shape is very close to the optimal shape. This is no longer true for larger optical depths where the efficiency can be 5 to $10\%$ higher by optimizing the input pulse shape. \\

\section{Implementation in rare-earth doped solids\label{implementation}}
We consider an implementation of the protocol in rare-earth-ion-doped crystals. Since this protocol is based on using the entire optical inhomogeneous linewidth, we require large ground-state splitting between $|g\rangle$ and $|s\rangle$  in order to obtain spectrally resolved transitions. This can be obtained with ions possessing Kramers Zeeman doublet states, such as Erbium and Neodymium ions. Indeed, Kramers ions have effective S=1/2 spins, resulting in a doublet with splitting $\Delta E = g \mu _B B$, where $g$ is the effective gyromagnetic factor, $B$ is the external magnetic field and $\mu _B \approx$ 14 GHz/Tesla the Bohr magneton. The $g$ factor is usually in the range 1-10, yielding splittings in the range 14-140 GHz/Tesla. Moreover we require high absorption coefficients in order to obtain an optical depth of around $\alpha L$ = 50, yielding efficiencies above 90\%.

We consider the system Nd:YVO$_4$, which has a transition at 879.705 nm ($^4I_{9/2} - ^4F_{3/2}$) with one of the highest oscillator strengths known for RE doped crystals \cite{Sun2002,McAuslan2009}. For a Neodymium doping concentration of 0.001\%, the absorption coefficient is $\alpha=41$ cm$^{-1}$ and the optical inhomogeneous broadening is $2\Gamma= 2\pi \cdot 2.1$ rad$\cdot$GHz \cite{Hastings-Simon2008}. The $g$ tensor has principal values of 2.36 and 0.915 \cite{Mehta2000}. If we assume a one centimetre long crystal ($L=1$ cm) and a Gaussian inhomogeneous spectrum, we find that the storage and retrieval efficiencies can be as high as 97\% when optimized over the input pulse shape (cf. Fig. \ref{fig6}). If we only optimize the duration of a pulse with Lorentzian spectrum (exponentially rising pulse in time) we find a slightly lower efficiency of 93\%, for a pulse duration given by the optimized condition $\Gamma T=6.25 \cdot 10^{-3}$. Consequently such a memory would be optimized for a pulse with an exponential rise time of $T/2=4.7$ ps in terms of intensity (see Eq. \ref{equInCon}). To estimate the required pulse energies for achieving efficient $\pi$ pulses, we need to assume a $\pi$ pulse duration $\tau$. We here consider a $\pi$ pulse 10 times shorter than the storage pulse, hence $\tau =$ 470 fs. If the $\pi$ pulse is too long as compared to the characteristic emission duration $T$, a substantial fraction of the stored light will immediately be emitted, this loss should be of the order $\tau/T$. Using the known dipole moment of Nd:YVO$_4$ and assuming a focused beam diameter of 50 $\mu m$ (with a Rayleigh range of 6 mm), we estimate the required $\pi$ pulse energy to be $E=600 \mu$J for $\tau=$ 470 fs. This pulse energy is rather high, indeed current fs-ps systems are able to deliver 10-100 $\mu$J at 883 nm. To increase the intensity, hence reducing pulse energies, one should focus tighter, which in turn requires a shorter crystal. The crystal length could be reduced substantially by working with higher concentration (100 ppm or above) and putting the crystal in a double-pass configuration (see for instance Ref. \cite{Clausen2011}). This would allow for an order of magnitude reduction in pulse energy. We also note that to achieve a 80\% efficiency one would only need an optical depth of $\alpha L=17$ for which the optimal condition is $\Gamma T=0.12$, resulting in an optimal input duration $T/2=9.1$ ps. The shorter input pulse and the shorter crystal ($L=2.1$ mm in double pass) where the focus can be tighter would cut the required pulse energy by roughly a factor of 8. We therefore believe it is realistic to achieve high efficiency and high bandwidth operation with this approach.

\section{Conclusion\label{Conclusion}}
We have analyzed a simple memory protocol which uses the natural inhomogeneous broadening inherently present in solid state systems.  The central idea behind the proposed protocol is to benefit from the free induction decay type of reemission arising when one stores pulses with a bandwidth larger than the inhomogeneous bandwidth. Coherent free induction decay is usually observed in the regime where the optical depth is of the order of unity. In that case, the emission only carries away a small fraction of the stored energy and the emission process is inefficient (a few percent). We here focused on the opposite regime of large optical depth ($ \alpha L \gg 1$ ) and showed that in this case, close to all the stored energy is carried away by the free induction emission, leading to an efficient memory, provided that the input pulse shape is an exponentially raising pulse in time with a duration of the order of $ (\alpha L\Gamma)^{-1}$. The basic principle of efficient free induction emission has been demonstrated in a recent experiment by Walther et al. \cite{Walther09}. They reported on the observation of free induction type emissions with efficiencies around 20\% in a Pr$^{3+}$-doped Y$_2$SiO$_5$ crystal, in the regime of moderate optical depths. It should be noted though that they worked with a spectrally narrow ensemble created via advanced optical pumping schemes, while our idea is to use the entire inhomogeneous broadening. To obtain a memory with on-demand readout and a long storage time, we propose to insert optical control pulses that transfer the coherence back and forth to a spin state.  We have investigated a potential implementation in Nd$^{3+}$:YVO$_4$ where high bandwidth and high storage efficiency could potentially be obtained. The main challenge in this system is to realize an efficient population transfer to the spin transition in order to make the readout on-demand and to achieve long storage times. In general, it will be easier to achieve an efficient transfer in an atomic ensemble exhibiting the same optical depth but with a narrower inhomogeneous line. Atomic vapours deserves attention in this framework. Finally, we believe that this memory protocol will be useful in experiments where high-bandwidth and high-rate memories are useful, such as in linear-optical quantum computing. It could also form the basic scheme for creating a microwave-to-optical interface using RE crystals coupled to superconducting devices.

\section{Acknowledgements} This work was supported by the EU project Qurep and by the NCCR project QSIT. We thank A. S$\o$rensen for fruitful discussions.

\section{References}

\bibliographystyle{unsrt}
\bibliography{qmcommon}

\end{document}